\documentclass[english,aps, prl]{revtex4}
\usepackage[T1]{fontenc}
\usepackage[latin9]{inputenc}
\usepackage{amsmath}
\usepackage{amssymb}
\usepackage{graphicx}
\usepackage{esint}

\makeatletter

\providecommand{\tabularnewline}{\\}

\@ifundefined{textcolor}{}
{%
 \definecolor{BLACK}{gray}{0}
 \definecolor{WHITE}{gray}{1}
 \definecolor{RED}{rgb}{1,0,0}
 \definecolor{GREEN}{rgb}{0,1,0}
 \definecolor{BLUE}{rgb}{0,0,1}
 \definecolor{CYAN}{cmyk}{1,0,0,0}
 \definecolor{MAGENTA}{cmyk}{0,1,0,0}
 \definecolor{YELLOW}{cmyk}{0,0,1,0}
}

\makeatother

\usepackage{babel}
\begin{document}

\title{Anisotropic charged impurity-limited carrier mobility in monolayer
phosphorene}

\author{Zhun-Yong Ong}

\affiliation{Institute of High Performance Computing, A{*}STAR, Singapore 138632}

\author{Gang Zhang}

\affiliation{Institute of High Performance Computing, A{*}STAR, Singapore 138632}

\author{Yong Wei Zhang}

\affiliation{Institute of High Performance Computing, A{*}STAR, Singapore 138632}
\begin{abstract}
The room temperature carrier mobility in atomically thin 2D materials
is usually far below the intrinsic limit imposed by phonon scattering
as a result of scattering by remote charged impurities in its environment.
We simulate the charged impurity-limited carrier mobility $\mu$ in
bare and encapsulated monolayer phosphorene. We find a significant
temperature dependence in the carrier mobilities ($\mu\propto T^{-\gamma}$)
that results from the temperature variability of the charge screening
and varies with the crystal orientation. The anisotropy in the effective
mass leads to an anisotropic carrier mobility, with the mobility in
the armchair direction about one order of magnitude larger than in
the zigzag direction. In particular, this mobility anisotropy is enhanced
at low temperatures and high carrier densities. Under encapsulation
with a high-$\kappa$ overlayer, the mobility increases by up to an
order of magnitude although its temperature dependence and its anisotropy
are reduced.

\end{abstract}
\maketitle
The search for alternatives to graphene for nanoelectronic applications
has expanded lately to include transition metal dichalcogenides (TMDs)\ \cite{BRadisavljevic:Nature11,QHWang:NatureNano12,HQiu:APL12_Electrical,DJariwala:ACSNano14_Emerging}
and other atomically thin two-dimensional (2D) crystals.\ \cite{HOChurchill:NatNanotech14,ESReich:Nature14_Phosphorene}
Phosphorene, an ultrathin form of black phosphorus (BP), has recently
garnered considerable interest because of its potentially high carrier
mobility ($\sim300$ cm$^{2}$V$^{-1}$s$^{-1}$ in few-layer samples\ \cite{SPKoenig:APL14_Electric}),
direct band gap\ \cite{HLiu:ACSNano14_Phosphorene} and electrical
conductance anisotropy.\ \cite{FXia:NatCommun14_Rediscovering} In
contrast, measurements of unprocessed monolayer TMD crystals have
yielded room-temperature mobilities typically below $10$ cm$^{2}$V$^{-1}$s$^{-1}$\ \cite{BBaugher:NL13}
although the mobility in similarly thick multilayer MoS$_{2}$ has
been measured to be around $200$ cm$^{2}$V$^{-1}$s$^{-1}$.\ \cite{SKim:NatComm11_High}
However, recent experimental studies of the hole mobility in few-layer
BP suggest that thinning phosphorene leads to a substantial reduction
in the mobility,\ \cite{LLi:NatNanotech14_Black,HLiu:ACSNano14_Phosphorene}
possibly due to the closer proximity between the charge carriers and
remote Coulomb impurities in the substrate. Extrapolating to a single
phosphorene layer, the carrier mobility would ultimately be limited
by charged impurity scattering even at room temperature, as in monolayer
MoS$_{2}$.\ \cite{ZYOng:PRB13_Mobility} 

Despite intense theoretical interest in monolayer phosphorene,\ \cite{ASRodin:PRL14_Strain,ANRudenko:PRB14_Quasiparticle}
there has not been a successful demonstration of a \emph{working}
monolayer phosphorene-based field-effect transistor (FET) to date.\ \cite{HLiu:ACSNano14_Phosphorene}
Nonetheless, the eventual realization of such a device is highly probable
in our opinion since monolayer phosphorene has been physically isolated\ \cite{HLiu:ACSNano14_Phosphorene}.
The atomic thinness of a monolayer 2D crystal also allows for higher
on-off current ratios, providing superior electrostatic modulation
of the channel carrier density via an external gate. In a FET, this
results in small off-currents and large switching ratios which are
advantageous for low-power device applications. Thus, it would be
advantageous to have a model of charge transport in \emph{supported}
monolayer phosphorene that takes into account its \emph{anisotropic}
character and can be used to interpret electrical transport data from
realistic phosphorene-based devices. Furthermore, understanding the
effect of the dielectric environment on charge transport provides
a basis for design strategies to optimize phosphorene-based device
performance.

\begin{figure}[p]
\includegraphics[width=12cm]{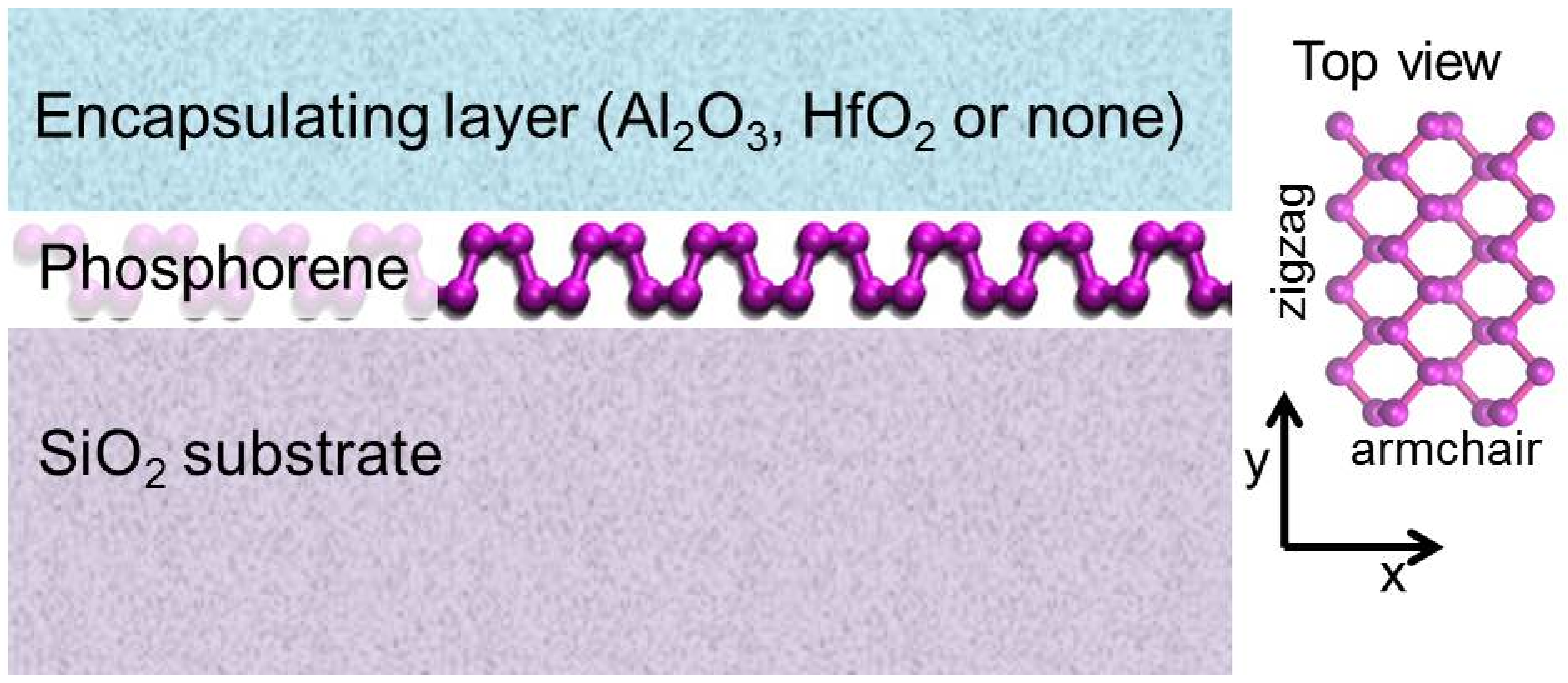}

\caption{Schematic of the supported phosphorene device. The side view shows
phosphorene in the armchair direction. The monolayer phosphorene is
sandwiched between a SiO$_{2}$ ($\kappa=3.9$) substrate and an encapsulating
layer which may be Al$_{2}$O$_{3}$ ($\kappa=12.5$), HfO$_{2}$
($\kappa=22$) or nothing ($\kappa=1$) for bare phosphorene. In the
top view of phosphorene on the right, we define the $x$-axis ($y$-axis)
to be parallel to the armchair (zigzag) direction, following the convention
in Refs.\ \cite{ANRudenko:PRB14_Quasiparticle,TLow:Arxiv14_Tunable}}

\label{Fig:DeviceConfiguration}
\end{figure}

In this paper, we study the charged impurity-limited electron and
hole transport in supported monolayer phosphorene as shown in Fig.\ \ref{Fig:DeviceConfiguration}.
We examine the the dependence of the drift mobility on orientation,
carrier type (electron vs. hole), temperature and dielectric environment.
Given the large anisotropy in effective mass, we study its effect
on the mobility anisotropy and how the anisotropy varies with carrier
density, temperature and dielectric screening. Hopefully, we provide
enough details of our model and results so that meaningful guidance
for and comparison with any future mobility measurements can be made
should a working monolayer phosphorene-based FET be made. In particular,
we focus on the anisotropy of the mobility and its temperature dependence
which in monolayer MoS$_{2}$ FETs has been shown to be caused by
the temperature variability of the charge screening\ \cite{ZYOng:PRB13_Mobility}
and is sometimes attributed to the inelastic phonon scattering of
electrons.\ \cite{DJariwala:APL13,BRadisavljevic:NatMat13} Given
that encapsulation with a high-$\kappa$ oxide insulator has been
used to enhance carrier mobility in ultrathin 2D crystals,\ \cite{BRadisavljevic:NatMat13}
we also model its effects on charge transport.

We state the main assumptions of our charge transport model. As in
monolayer graphene and TMD crystals, we suppose that charge transport
in the \emph{metallic} phase\ \cite{BRadisavljevic:NatMat13} in
monolayer phosphorene is dominated by charged impurity (CI) scattering
at low and room temperature.\ \cite{ZYOng:PRB13_Mobility} The calculated
intrinsic phonon-limited carrier mobility in monolayer graphene\ \cite{RSShishir:JPhys09}
and TMD crystals\ \cite{KKaasbjerg:PRB12,XLi:PRB13_IntrinsicMoS2}
is usually much higher than what is measured in experiments,\ \cite{YWTan:PRL07,SGhatak:ACSNano11,BBaugher:NL13,BRadisavljevic:NatMat13,DJariwala:ACSNano14_Emerging,DOvchinnikov:ACSNano14_Electrical,ZYu:NatCommun14}
suggesting that defect and charged impurity scattering are the key
mobility-limiting mechanism even at room temperature. For example,
the phonon-limited electron mobility in single-layer MoS$_{2}$ is
estimated\ \cite{KKaasbjerg:PRB12,NMa:PRX14_Charge} to be around
200 to 410 cm$^{2}$V$^{-1}$s$^{-1}$ at room temperature while the
measured electron mobility is at least one order of magnitude smaller
at around 20 cm$^{2}$V$^{-1}$s$^{-1}$ even in the metallic phase.\ \cite{BBaugher:NL13,BRadisavljevic:NatMat13}
Furthermore, the measured mobility in single-layer TMD crystals also
exhibit\ \cite{SGhatak:ACSNano11,DOvchinnikov:ACSNano14_Electrical}
a strong carrier density dependence as predicted by charged impurity-limited
transport models.\cite{ZYOng:PRB13_Mobility} Although the electrical
transport properties of monolayer phosphorene have not been characterized,
independent measurements of the hole mobility\ \cite{HLiu:ACSNano14_Phosphorene,LLi:NatNanotech14_Black}
have shown that the room-temperature hole mobility in multi-layer
BP decreases sharply as the crystal thickness is progressively reduced.
This is probably due to the greater proximity of the active carrier
channel to remote charged impurities near the BP-substrate interface
or from physisorbed species. Thus, carrier scattering by phonons in
our calculations of the mobility is ignored, and we believe that the
temperature dependence of charge transport under existing experimental
conditions, where charged impurity densities are high, is largely
captured by carrier scattering by charged impurities with temperature-sensitive
polarization charge screening playing a crucial role.\ \cite{ZYOng:PRB13_Mobility,NMa:PRX14_Charge}
The effects of interface impurity traps\ \cite{WZhu:NatComm14_Electronic}
are also excluded. The charge transport model is developed within
the framework of the semi-classical Boltzmann transport theory, an
approach that has been successfully applied to understand charge transport
in few- and single-layer TMDs.\ \cite{SKim:NatComm11_High,ZYOng:PRB13_Mobility,NMa:PRX14_Charge}

We also assume that the electron dispersion in monolayer phosphorene
is ellipsoidal both in the conduction and valence band at the $\Gamma$-point
of the Brillouin zone, as in Ref.\ \cite{TLow:Arxiv14_Tunable} where
the effective masses are extracted from a low-energy effective Hamiltonian.\ \cite{ASRodin:PRL14_Strain}
In the conduction band, the electron energy at $\mathbf{k}=(k_{x},k_{y})$
is $E_{c}(\mathbf{k})=\frac{\hbar^{2}}{2}\left(\frac{k_{x}^{2}}{m_{x}^{(e)}}+\frac{k_{y}^{2}}{m_{y}^{(e)}}\right)+\frac{1}{2}E_{g}$
where $m_{x}^{(e)}$ and $m_{y}^{(e)}$ are the effective masses in
the $x$ and $y$ direction, respectively. $E_{g}$ is the band gap.
Likewise in the valence band, the electron energy is given by $E_{v}(\mathbf{k})=-\frac{\hbar^{2}}{2}\left(\frac{k_{x}^{2}}{m_{x}^{(h)}}+\frac{k_{y}^{2}}{m_{y}^{(h)}}\right)-\frac{1}{2}E_{g}$
where $m_{x}^{(h)}$ and $m_{y}^{(h)}$ are the effective masses in
the $x$ and $y$ direction, respectively. Table\ \ref{Tab:EffectiveMasses}
shows the effective mass values used. We follow the convention used
in Ref.\ \cite{TLow:Arxiv14_Tunable}: the $x$-direction ($y$-direction)
corresponds to the so-called armchair (zigzag) direction where we
have light (heavy) electrons and holes.

\section*{Methods}

\subsection*{Calculation of charged impurity mobility}

To avoid the tedium of repeating ourselves, we drop the superscript
labels for electrons and holes in the description of our charge transport
model. To obtain the desired electron- or hole-dominated quantity,
we replace the effective mass in the formula with the electron or
hole effective mass. The charge current density $\mathbf{j}$ is related
to the applied electric field via the expression $\mathbf{j}=\boldsymbol{\sigma}\cdot\mathbf{E}$
where $\boldsymbol{\sigma}$ is the 2D conductivity tensor. The non-zero
diagonal elements of $\boldsymbol{\sigma}$ are $\sigma_{xx}$ and
$\sigma_{yy}$. In the relaxation time approximation,\ \cite{CJacoboni:Book2010}
the formula for $\sigma_{xx}$ is 
\begin{equation}
\sigma_{xx}=\frac{ge^{2}}{(2\pi)^{2}k_{B}T}\int d\mathbf{k}\ \tau(\mathbf{k})v_{x}(\mathbf{k})^{2}f(\mathbf{k})[1-f(\mathbf{k})]\ ,\label{Eq:ConductivityTensor}
\end{equation}
where $\tau$ is the momentum relaxation time from CI scattering and
$v_{x}$ is the group velocity in the $x$-direction, given by $v_{x}=\hbar k_{x}/m_{x}$.
$g$, $k_{B}$, $T$, $e$ and $f$ are the spin degeneracy ($g=2$),
the Boltzmann constant, the temperature, the electron charge and the
equilibrium Fermi-Dirac distribution, respectively. $\sigma_{yy}$
is similarly defined as in Eq.\ (\ref{Eq:ConductivityTensor}). We
define the drift mobility in the armchair ($x$) and zigzag ($y$)
direction as $\mu_{\text{AC}}=\sigma_{xx}/n$ and $\mu_{\text{ZZ}}=\sigma_{yy}/n$,
respectively, where $n$ is the carrier density. As shown in Fig.\ \ref{Fig:DeviceConfiguration},
the phosphorene layer is sandwiched between a SiO$_{2}$ substrate
and a top encapsulating layer that can be either Al$_{2}$O$_{3}$,
HfO$_{2}$ or nothing in the case of bare phosphorene. The 2D integral
in Eq.\ (\ref{Eq:ConductivityTensor}) is evaluated numerically using
our own computer code.

The limiting variable that determines the conductance in Eq.\ (\ref{Eq:ConductivityTensor})
is the momentum relaxation time, which corresponds to the mean free
time between scattering events and is the inverse of the momentum
relaxation rate. Charge carrier momentum loss is caused by elastic
scattering with screened charged impurities at the phosphorene-SiO$_{2}$
interface. We assume that charged impurity is isotropic \emph{i.e.}
its orientation in 2D does not matter even though the effective masses
are highly anisotropic. We also assume that the CI concentration is
sufficiently dilute such that multiple scattering\ \cite{BKRidley:Book99}
can be ignored and that the positions of the impurities are uncorrelated.\ \cite{QLi:PRL11_Theory,RAnicic:PRB13_Effects}

We write the momentum relaxation rate $\Gamma(\mathbf{k})$ as\ \cite{CJacoboni:Book2010}
\begin{align}
\Gamma(\mathbf{k}) & =\frac{n_{\mathrm{imp}}}{2\pi\hbar}\int d\mathbf{k^{\prime}}|\phi(\mathbf{k},\mathbf{k^{\prime}})|^{2}\nonumber \\
 & \times\left[1-\frac{\mathbf{v}(\mathbf{k})\cdot\mathbf{v}(\mathbf{k^{\prime}})}{|\mathbf{v}(\mathbf{k})||\mathbf{v}(\mathbf{k^{\prime}})|}\right]\delta[E(\mathbf{k})-E(\mathbf{k^{\prime}})]\label{Eq:MomentumRelaxationRate}
\end{align}
where $n_{\mathrm{imp}}$ is the CI density concentration, and $\phi(\mathbf{k},\mathbf{k^{\prime}})$
is the scattering potential given by\ \cite{ZYOng:PRB12_TopGate}
\begin{equation}
\phi(\mathbf{k},\mathbf{k^{\prime}})=\frac{e^{2}G(\mathbf{k}-\mathbf{k^{\prime}})}{1-e^{2}G(\mathbf{k}-\mathbf{k^{\prime}})\Pi(\mathbf{k}-\mathbf{k^{\prime}},E_{F},T)}\ .\label{Eq:ScatteringPotential}
\end{equation}
$G(\mathbf{q})$ is the Green's function to the Poisson equation and
is equal to $G(\mathbf{q})=[(\epsilon_{\mathrm{box}}+\epsilon_{\mathrm{tox}})q]^{-1}$
where $\epsilon_{\mathrm{box}}$ and $\epsilon_{\mathrm{tox}}$ are
the permittivities of the oxide layers under and above the phosphorene.
For bare phosphorene, $\epsilon_{tox}=\epsilon_{0}$ where $\epsilon_{0}$
is the permittivity of vacuum. 

The static polarizability $\Pi(\mathbf{k}-\mathbf{k^{\prime}},T,E_{F})$
in Eq.\ (\ref{Eq:ScatteringPotential}) is isotropic at 0K.\ \cite{TLow:Arxiv14_Tunable}
To make the problem tractable, we assume that $\Pi(\mathbf{k}-\mathbf{k^{\prime}},T,E_{F})$
remains isotropic at finite temperatures and for all wavelengths.
This allows us to model the static polarizability as that of an isotropic
2D electron gas with an effective mass of $m_{\mathrm{eff}}=\sqrt{m_{x}m_{y}}$,
using the expression\ \cite{PMaldague:SurfSci1978,ZYOng:PRB13_Mobility,NMa:PRX14_Charge}
\begin{equation}
\Pi(q,T,E_{F})=\int_{0}^{\infty}d\mu\ \frac{\Pi(q,0,\mu)}{4k_{B}T\cosh(\frac{E_{F}-\mu}{2k_{B}T})}\label{Eq:Polarizability}
\end{equation}
where $\Pi(q,0,\mu)=-\frac{m_{\mathrm{eff}}}{\pi\hbar^{2}}\left[1-\Theta(q-2k_{F})\sqrt{1-(\frac{2k_{F}}{q})^{2}}\right]$
and $k_{F}=\sqrt{2m_{\mathrm{eff}}\mu}/\hbar$. $E_{F}$ is the chemical
potential and is determined by $E_{F}=k_{B}T\ln\left\{ \exp[\pi\hbar^{2}n/(m_{\textrm{eff}}k_{B}T)]-1\right\} $.

\section*{Results}

\begin{table}[p]
\begin{tabular}{|c|c|c|}
\hline 
 & Hole & Electron\tabularnewline
\hline 
\hline 
Armchair ($x$-direction)  & $m_{x}^{(h)}=0.15m_{0}$ & $m_{x}^{(e)}=0.15m_{0}$\tabularnewline
\hline 
Zigzag ($y$-direction) & $m_{y}^{(h)}=0.7m_{0}$ & $m_{y}^{(e)}=1.0m_{0}$\tabularnewline
\hline 
\end{tabular}

\caption{Electron and hole effective mass values used in our calculations.
The parameters are taken from Refs.\ \ \cite{ASRodin:PRL14_Strain,TLow:Arxiv14_Tunable}.
$m_{0}$ is the mass of the free electron.}

\label{Tab:EffectiveMasses}
\end{table}

\subsection*{Electron and hole mobility in bare unencapsulated phosphorene}

\begin{figure}[p]

\includegraphics[width=13.5cm]{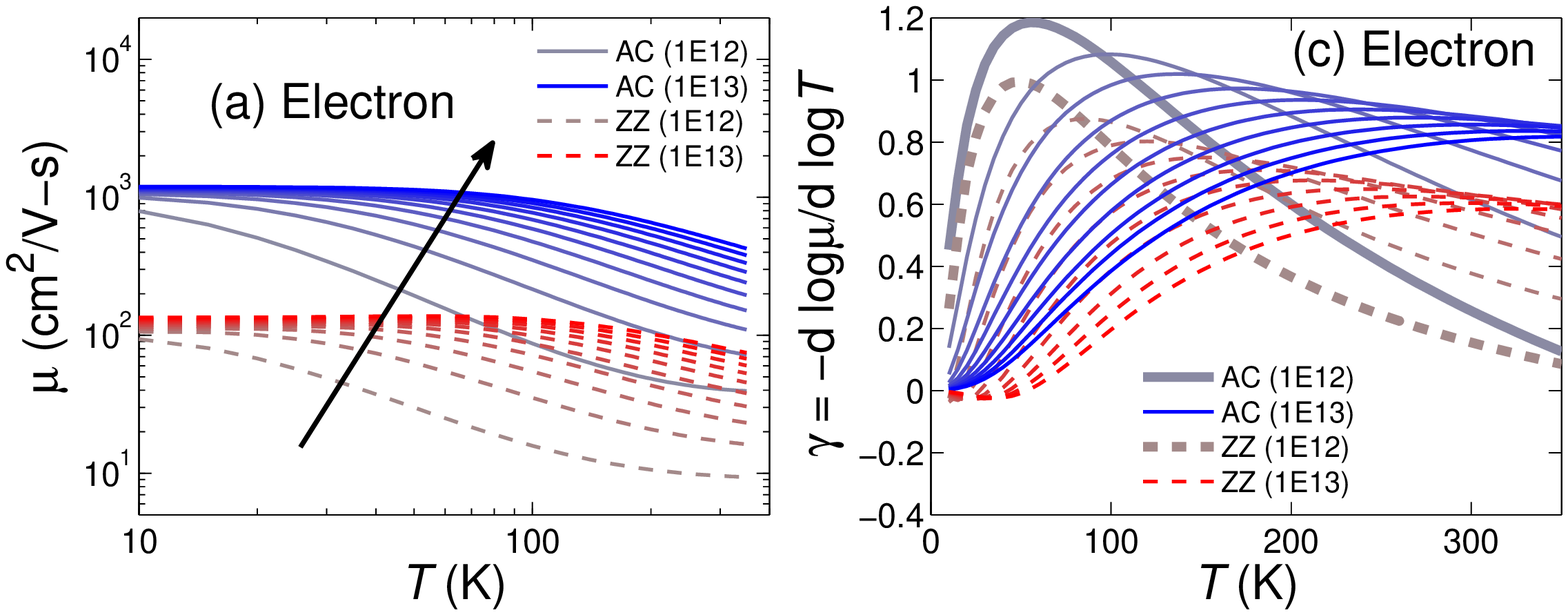}

\includegraphics[width=13.5cm]{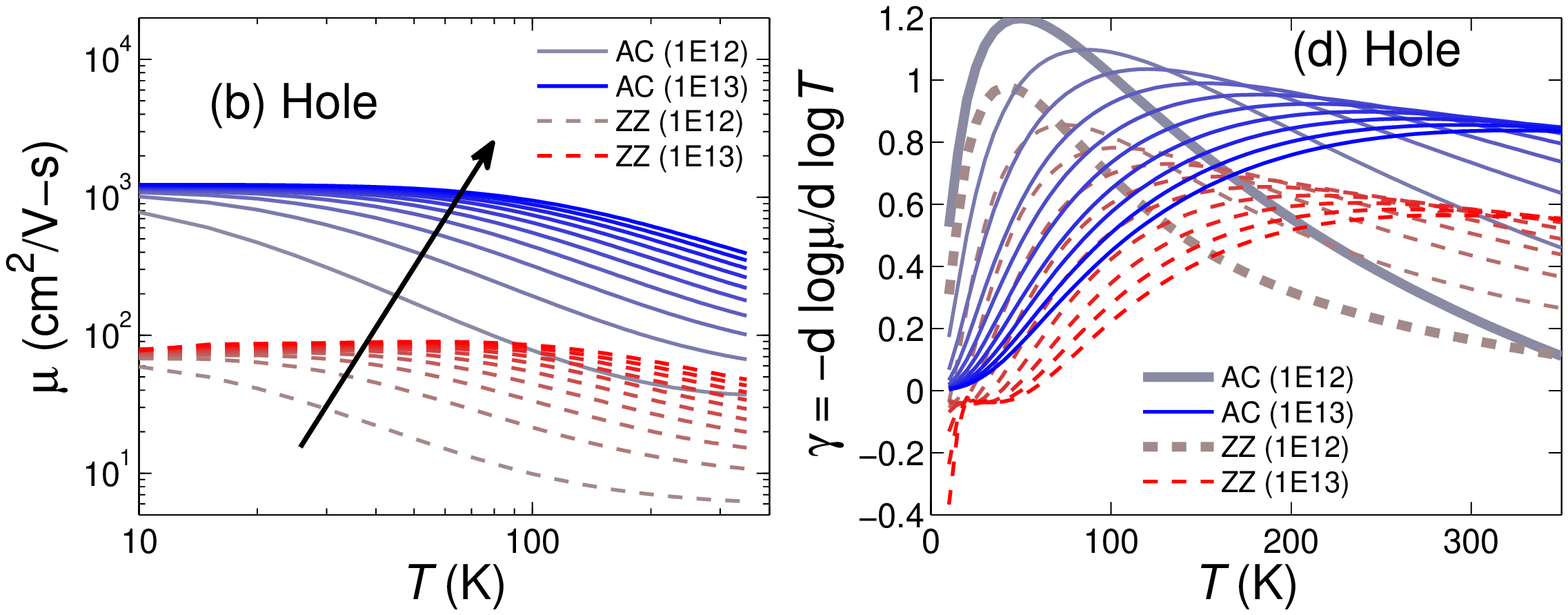}\caption{Plot of the CI-limited \textbf{(a)} electron and \textbf{(b)} hole
mobility $\mu$ in the armchair (solid line) and zigzag (dashed line)
directions from $n=10^{12}$ to $10^{13}$ cm$^{-2}$ in steps of
$\Delta n=10^{12}$ cm$^{-2}$ where $n$ is the carrier (electron
or hole) density, for unencapsulated phosphorene, assuming a charged
impurity density of $n_{\text{imp}}=10^{12}$ cm$^{-2}$. We label
low-density ($n=10^{12}$ cm$^{-2}$) mobility in the armchair and
zigzag direction as `AC (1E12)' and `ZZ (1E12)',respectively. The
corresponding high-density ($n=10^{13}$ cm$^{-2}$) mobilities are
labeled `AC (1E13)' and `ZZ (1E13)'. The black arrows indicate the
direction of increasing carrier density. Since $m_{x}^{(e)}<m_{y}^{(e)}$
and $m_{x}^{(h)}<m_{y}^{(h)}$ , the mobilities in the armchair ($x$)
direction $\mu_{\text{AC}}$ are about one order of magnitude larger
than the mobilities in the zigzag ($y$) direction $\mu_{\text{ZZ}}$.
The corresponding power law exponents {[}$\gamma=-d(\log\mu)/d(\log T)${]}
for \textbf{(c)} electrons and \textbf{(d)} holes are shown as a function
of temperature. At room temperature (300 K), $\gamma$ is higher for
$\mu_{\text{AC}}$ than for $\mu_{\text{ZZ}}$ for both electrons
and holes \emph{i.e.} $\mu_{\text{AC}}$ is more temperature sensitive.}

\label{Fig:MobilityVsTemperature_bare}
\end{figure}

We compute the CI-limited drift mobilities $\mu_{\text{AC}}$ and
$\mu_{\text{ZZ}}$ for electrons and holes in \emph{bare} monolayer
phosphorene (Fig.\ \ref{Fig:MobilityVsTemperature_bare}), assuming
an impurity density of $n_{\textrm{imp}}=10^{12}$ cm$^{-2}$ which
is typical of the SiO$_{2}$ substrate used in experiments.\ \cite{KMBurson:NanoLett13_Direct}
Note that $n_{\textrm{imp}}$ can be allowed to freely vary and that
the CI-limited mobility is inversely proportional to $n_{\textrm{imp}}$
\emph{i.e.} $\mu\propto1/n_{\textrm{imp}}$ since the scattering rate
is proportional to $n_{\textrm{imp}}$ {[}see Eq.\ (\ref{Eq:MomentumRelaxationRate}){]}.
Hence, our mobility results are not fixed but only \emph{representative}
values since the mobility can always be rescaled by varying the impurity
density $n_{\text{imp}}$ although the temperature dependence does
not vary with $n_{\textrm{imp}}$.

Figure\ \ref{Fig:MobilityVsTemperature_bare}(a-b) shows $\mu_{\text{AC}}$
and $\mu_{\text{ZZ}}$ for electrons and holes in bare phosphorene.
As expected, the mobility is about one order of magnitude larger in
the armchair direction than in the zigzag direction for both electrons
and holes because of the much larger effective masses in the zigzag
direction. The electron and hole $\mu_{\text{AC}}$ values are close
because $m_{x}^{(e)}=m_{x}^{(h)}$. On the other hand, in the zigzag
($y$) direction, the electron mobility is higher than the hole mobility
because $m_{y}^{(e)}<m_{y}^{(h)}$. We also observe that the mobility
decreases with temperature for all carrier types and crystal orientations
(armchair and zigzag). Like in monolayer MoS$_{2}$,\ \cite{ZYOng:PRB13_Mobility}
the mobility drop is carrier density-dependent with the inflection
at a lower temperature for lower carrier densities. At any given temperature,
the mobility increases monotonically with carrier density, and at
room temperature, it varies almost linearly with carrier density.
At 300 K, the mobility in cm$^{2}$V$^{-1}$s$^{-1}$ can be fitted
to the following empirical formulae\begin{subequations}

\begin{eqnarray}
\mu_{\text{AC}} & = & \frac{1}{n_{\text{imp}}}\times\begin{cases}
45.9n-33.7+\frac{42.9}{1+n^{2}} & \text{for electrons}\\
42.5n-33.5+\frac{42.3}{1+n^{2}} & \text{for holes}
\end{cases}\label{Eq:EmpiricalFit_AC}\\
\mu_{\text{ZZ}} & = & \frac{1}{n_{\text{imp}}}\times\begin{cases}
7.32n+1.57 & \text{for electrons}\\
4.63n+1.49 & \text{for holes}
\end{cases}\label{Eq:EmpiricalFit_ZZ}
\end{eqnarray}
\label{Eq:EmpiricalFits}\end{subequations} where $n_{\text{imp}}$
and $n$ are the CI impurity and carrier concentration in $10^{12}$
cm$^{-2}$, respectively. The first term in Eq.\ (\ref{Eq:EmpiricalFit_AC})
represents the linear $n$-dependence of the charged impurity-limited
carrier mobility that is typical of the 2-dimensional electron gas
with parabolic dispersion and weak screening.\ \cite{SAdam:PRB08_Boltzmann,SGhatak:ACSNano11}
The second and third terms in Eq.\ (\ref{Eq:EmpiricalFit_AC}) are
corrections taking into account the deviation from the purely linear
behavior in lightly and heavily doped samples. Similarly, the mobility
in the zigzag direction in Eq.\ (\ref{Eq:EmpiricalFit_AC}) can be
expressed as the sum of a term linearly dependent on $n$ and a density-independent
term which yields the mobility in undoped phosphorene.

It has been claimed\ \cite{LLi:NatNanotech14_Black,FXia:NatCommun14_Rediscovering}
that the phosphorene hole mobility follows a power law \emph{i.e.}
$\mu\propto T^{-\gamma}$ at $T>100$ K because of inelastic phonon
scattering.\ \cite{AMorita:JPSP89_Electron} To facilitate the analysis
of the temperature dependence, we compute the temperature-dependent
exponent $\gamma$ by taking numerically the negative logarithmic
derivative of the mobility with respect to $\log T$, \emph{i.e.}
$\gamma=-d(\log\mu)/d(\log T)$, which at high temperatures ($T>100$
K) can be interpreted as the power law exponent if the mobility scales
as $\mu\propto T^{-\gamma}$. The results are shown in Fig.\ \ref{Fig:MobilityVsTemperature_bare}(c)
and (d) for electron and holes, respectively. We observe that $\gamma$
is significantly larger for $\mu_{\text{AC}}$ than for $\mu_{\text{ZZ}}$
for both electrons and holes. At $n=10^{13}$ cm$^{-2}$ and $T=350$
K, $\gamma\sim0.8$ ($0.6$) in the armchair (zigzag) direction for
both electrons and holes, with $\gamma$ being slightly larger for
electrons. Thus, $\mu_{\text{AC}}$ is more temperature sensitive
than $\mu_{\text{ZZ}}$. The relatively large room temperature values
of $\gamma$ also indicate a significant temperature dependence even
in the absence of inelastic phonon scattering. Thus, due caution must
be exercised in interpreting the temperature dependence of mobility
values from experiments. Electrical characterization of few-layer
phosphorene FETs\ \cite{LLi:NatNanotech14_Black} gives $\gamma\sim0.5$
which has been construed as a signature of inelastic electron-phonon
scattering. However, our result for monolayer phosphorene raises the
possibility that CI scattering can play a significant role in the
mobility temperature dependence.

\subsection*{Effect of high-$\kappa$ oxide encapsulation}

Experiments have shown that the use of a high-$\kappa$ encapsulating
overlayer such as HfO$_{2}$ can enhance the electron mobility in
monolayer MoS$_{2}$ as well as weakens its temperature dependence.\ \cite{BRadisavljevic:NatMat13}
The reason for this mobility enhancement and weaker temperature dependence
has been attributed to the dielectric screening of the remote charge
impurities by the high-$\kappa$ top oxide.\ \cite{ZYOng:PRB13_Mobility,NMa:PRX14_Charge}
We explore the effect of high-$\kappa$ oxide encapsulating overlayers
such as Al$_{2}$O$_{3}$ ($\kappa=12.5$) and HfO$_{2}$ ($\kappa=22$)
on the CI-limited mobility in phosphorene.

\begin{figure}[p]

\includegraphics[width=13.5cm]{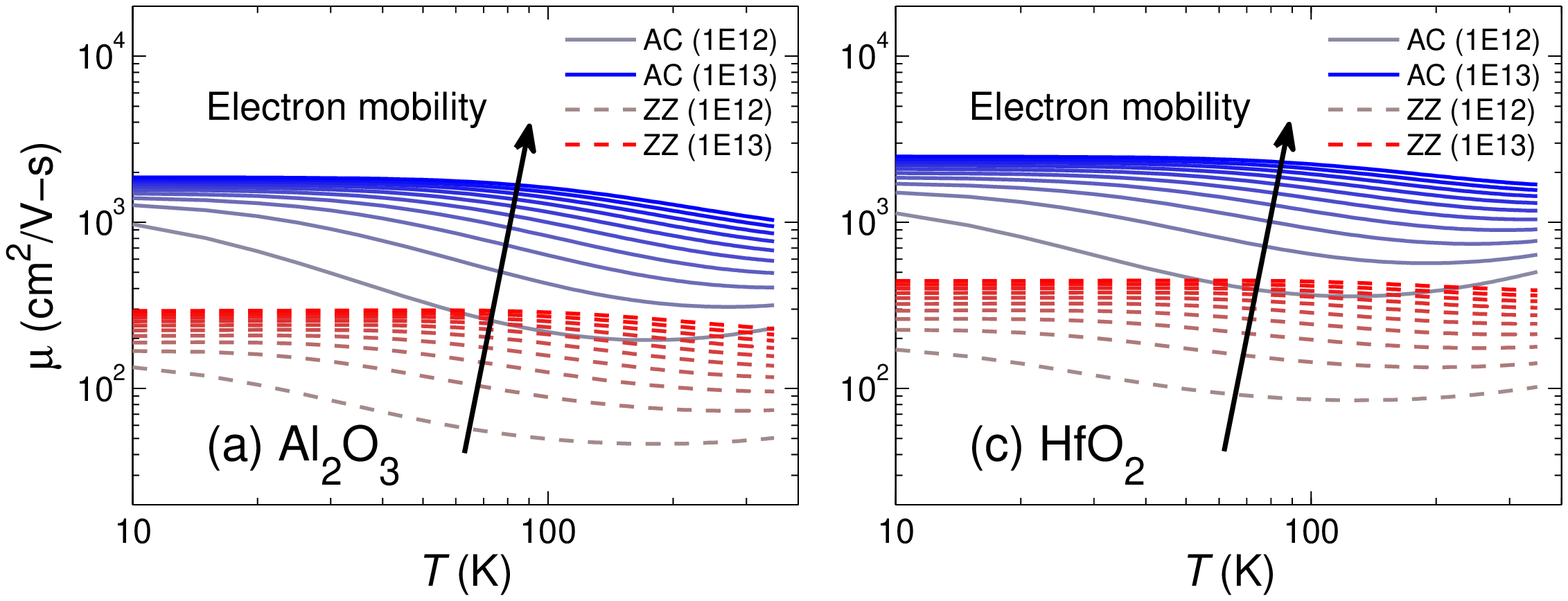}

\includegraphics[width=13.5cm]{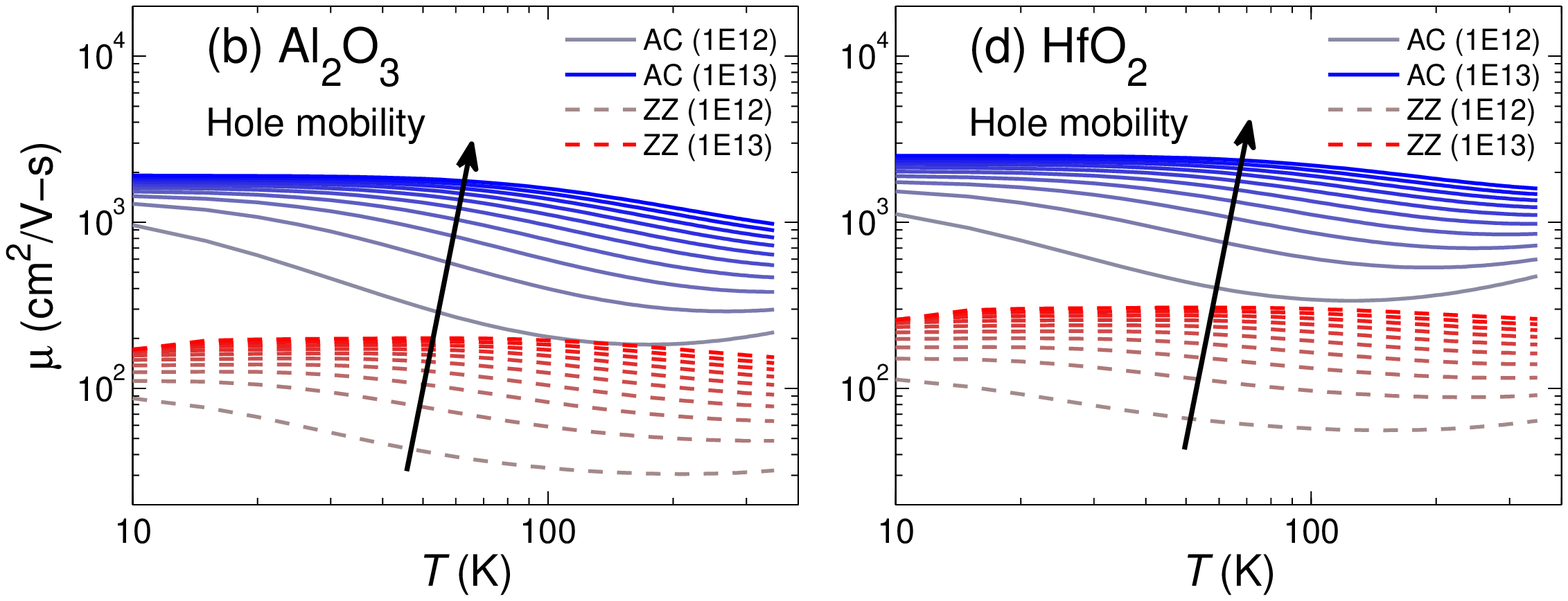}

\caption{Plot of the CI-limited mobility in the armchair and zigzag direction
from $n=10^{12}$ to $10^{13}$ cm$^{-2}$ in steps of $\Delta n=10^{12}$
cm$^{-2}$ for \textbf{(a)} electrons and \textbf{(b)} holes in Al$_{2}$O$_{3}$-covered
phosphorene as well as for \textbf{(c)} electrons and \textbf{(d)}
holes in HfO$_{2}$-covered phosphorene for $T=10$ to $350$K, assuming
a charged impurity density of $n_{\text{imp}}=10^{12}$ cm$^{-2}$.
Encapsulation reduces the mobility temperature dependence. At low
$n$, the electron mobility \emph{increases} with temperature.}

\label{Fig:HighKappaMobilities}
\end{figure}

Figure\ \ref{Fig:HighKappaMobilities} shows $\mu_{\text{AC}}$ and
$\mu_{\text{ZZ}}$ for electrons and holes in Al$_{2}$O$_{3}$- and
HfO$_{2}$-covered phosphorene, assuming a CI concentration of $n_{\text{imp}}=10^{12}$
cm$^{-2}$. Comparing Fig.\ \ref{Fig:HighKappaMobilities}(a) and
(b) with Fig.\ \ref{Fig:MobilityVsTemperature_bare}(a) and (b),
the electron and hole mobilities in encapsulated phosphorene are substantially
higher because the stronger dielectric screening of the charged impurities
by the high-$\kappa$ overlayer reduces their effective charge and
leads to less CI scattering. The mobility enhancement is even higher
for HfO$_{2}$ {[}see Fig.\ \ref{Fig:HighKappaMobilities}(c) and
(d){]} because its dielectric constant is larger ($\kappa=22$) and
screens the charged impurities more effectively. This mobility enhancement
is especially large at high temperatures and low carrier densities
where screening by the polarization charge is weak. For example, from
Fig.\ \ref{Fig:MobilityVsTemperature_bare}(a) and Fig.\ \ref{Fig:HighKappaMobilities}(a)
and (c), we find that $\mu_{\text{AC}}$= 41, 217 and 461 cm$^{2}$V$^{-1}$s$^{-1}$
at $n=10^{12}$ cm$^{-2}$ and $T=300$ K for electrons in unencapsulated,
Al$_{2}$O$_{3}$-covered and HfO$_{2}$-covered phosphorene, respectively
\emph{i.e.} encapsulation enhances the mobility by an order of magnitude.
A comparison between Fig.\ \ref{Fig:MobilityVsTemperature_bare}(a-b)
and Fig.\ \ref{Fig:HighKappaMobilities} also shows that encapsulation
leads a proportionally greater increase in $\mu_{\text{ZZ}}$ than
in $\mu_{\text{AC}}$.

While the high dielectric environment reduces the effective charge
on the charged impurities, it also weakens their polarization charge
screening.\ \cite{ZYOng:PRB13_Mobility} The latter effect has been
used to explain\ \cite{ZYOng:PRB13_Mobility} the weaker electron
mobility temperature dependence in HfO$_{2}$-covered monolayer MoS$_{2}$
relative to bare MoS$_{2}$.\ \cite{BRadisavljevic:NatMat13} Since
the temperature dependence of the mobility in unencapsulated phosphorene
is due to the temperature variability of charge screening, we expect
the mobility temperature dependence in encapsulated phosphorene to
be similarly modified by the high-$\kappa$ oxide overlayer.

\begin{figure}[p]

\includegraphics[width=13.5cm]{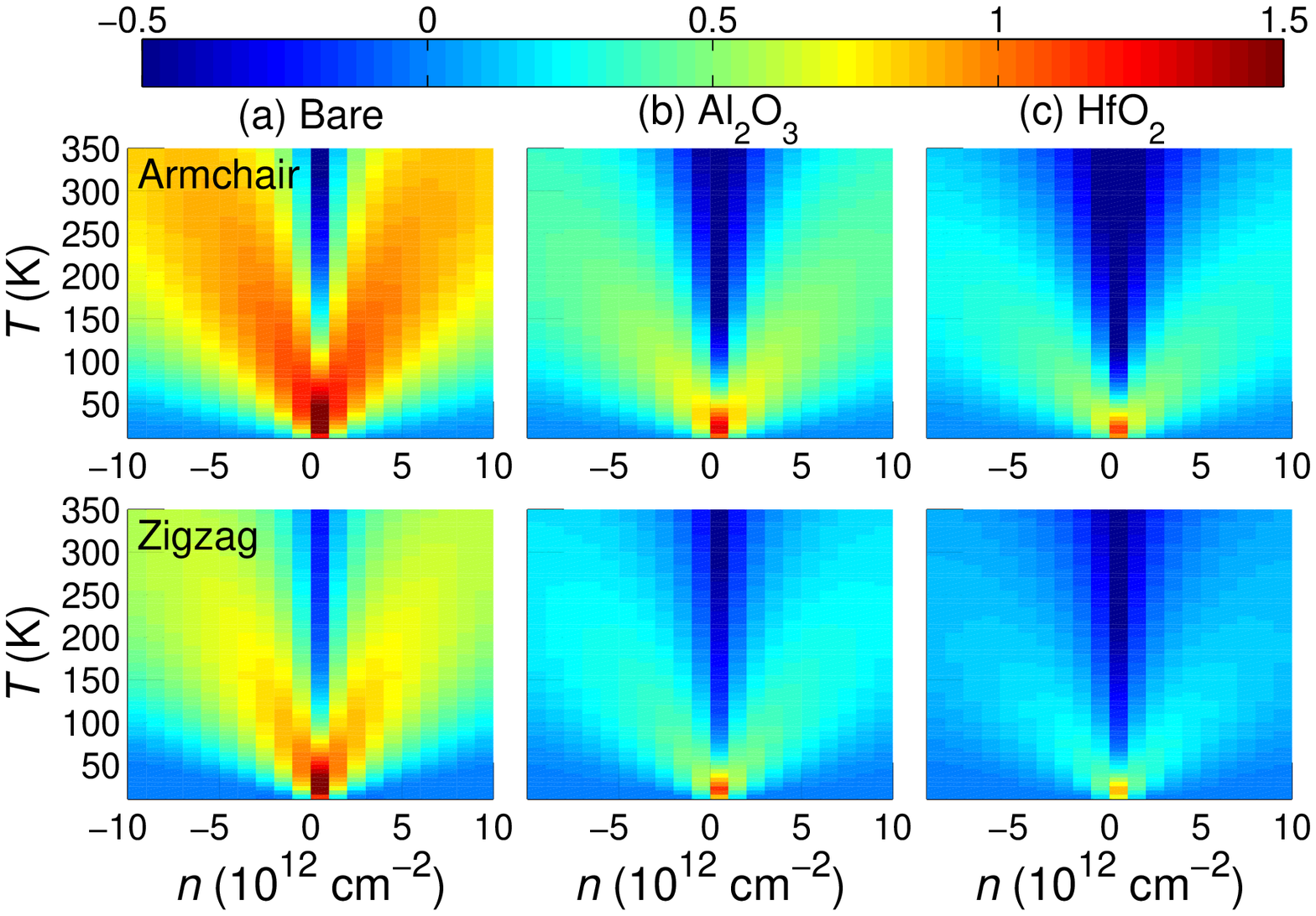}

\caption{We plot $\gamma$ {[}defined as $\gamma=-d(\log\mu)/d(\log T)${]}
for electrons and holes in phosphorene \textbf{(a)} without, \textbf{(b)}
with Al$_{2}$O$_{3}$ and \textbf{(c)} with HfO$_{2}$ encapsulationfor
$n=-10^{13}$ to $10^{13}$ cm$^{-2}$ from $T=10$ to 350 K. The
electron-doped (hole-doped) region corresponds to positive (negative)
$n$. The top and bottom panels correspond to $\gamma$ for $\mu_{\text{AC}}$
and $\mu_{\text{ZZ}}$, respectively. $\gamma$ is almost symmetric
about the neutral point ($n=0$). We find that magnitude of $\gamma$
is larger for $\mu_{\text{AC}}$ than for $\mu_{\text{ZZ}}$. Also,
$\gamma$ is reduced when phosphorene is encapsulated with Al$_{2}$O$_{3}$
or HfO$_{2}$, and at low densities and room temperature, becomes
negative \emph{i.e.} the mobility increases with temperature.}

\label{Fig:GammaAllOxides}
\end{figure}

For a systematic comparison of the temperature dependence, we show
the negative logarithmic derivative of the mobility with respect to
$\log T$ {[}\emph{i.e.} $\gamma=-d(\log\mu)/d(\log T)${]} as a function
of temperature and carrier density for $\mu_{\text{AC}}$ and$\mu_{\text{ZZ}}$
in Fig.\ \ref{Fig:GammaAllOxides}. A large $\gamma$ magnitude corresponds
to a strong mobility temperature dependence while a small magnitude
indicates a weak temperature dependence. Comparing Fig.\ \ref{Fig:GammaAllOxides}(a)
with Figs.\ \ref{Fig:GammaAllOxides}(b) and (c), we find that magnitude
of $\gamma$ for the phosphorene electrons and holes decreases after
encapsulation with Al$_{2}$O$_{3}$ or HfO$_{2}$, over the range
of carrier density and temperature values considered. This decrease
in the temperature dependence is due to the weakened polarization
charge screening. 

At low carrier densities ($n\sim10^{12}$ cm$^{-2}$), the electron
and hole $\gamma$ decreases and falls below zero as $T$ rises {[}see
Fig.\ \ref{Fig:GammaAllOxides}(b) and (c){]} \emph{i.e.} the mobility
actually increases with temperature for $T>200$ K, which can be most
clearly observed for the $\mu_{\text{AC}}$ curves corresponding to
$n=10^{12}$ cm$^{-2}$ in Fig.\ \ref{Fig:HighKappaMobilities}(c)
and (d). Physically, this is due to the weak polarization charge screening
at low carrier densities and high temperatures, which causes the scattering
cross section of the charged impurities to be effectively temperature
independent. Hence, as the temperature rises, more electron and holes
are thermally excited to higher energy states which have a smaller
CI scattering cross section and momentum relaxation rate, resulting
in a mobility increase. At high carrier densities ($n\sim10^{13}$
cm$^{-2}$), the electron and hole $\gamma$ is still positive with
the range of temperatures considered \emph{i.e.} the mobility still
falls as the temperature increases (see the $10^{13}$ cm$^{-2}$
$\mu_{\text{AC}}$ and $\mu_{\text{ZZ}}$ curves in Fig.\ \ref{Fig:HighKappaMobilities}).

\begin{figure}[p]

\includegraphics[width=13.5cm]{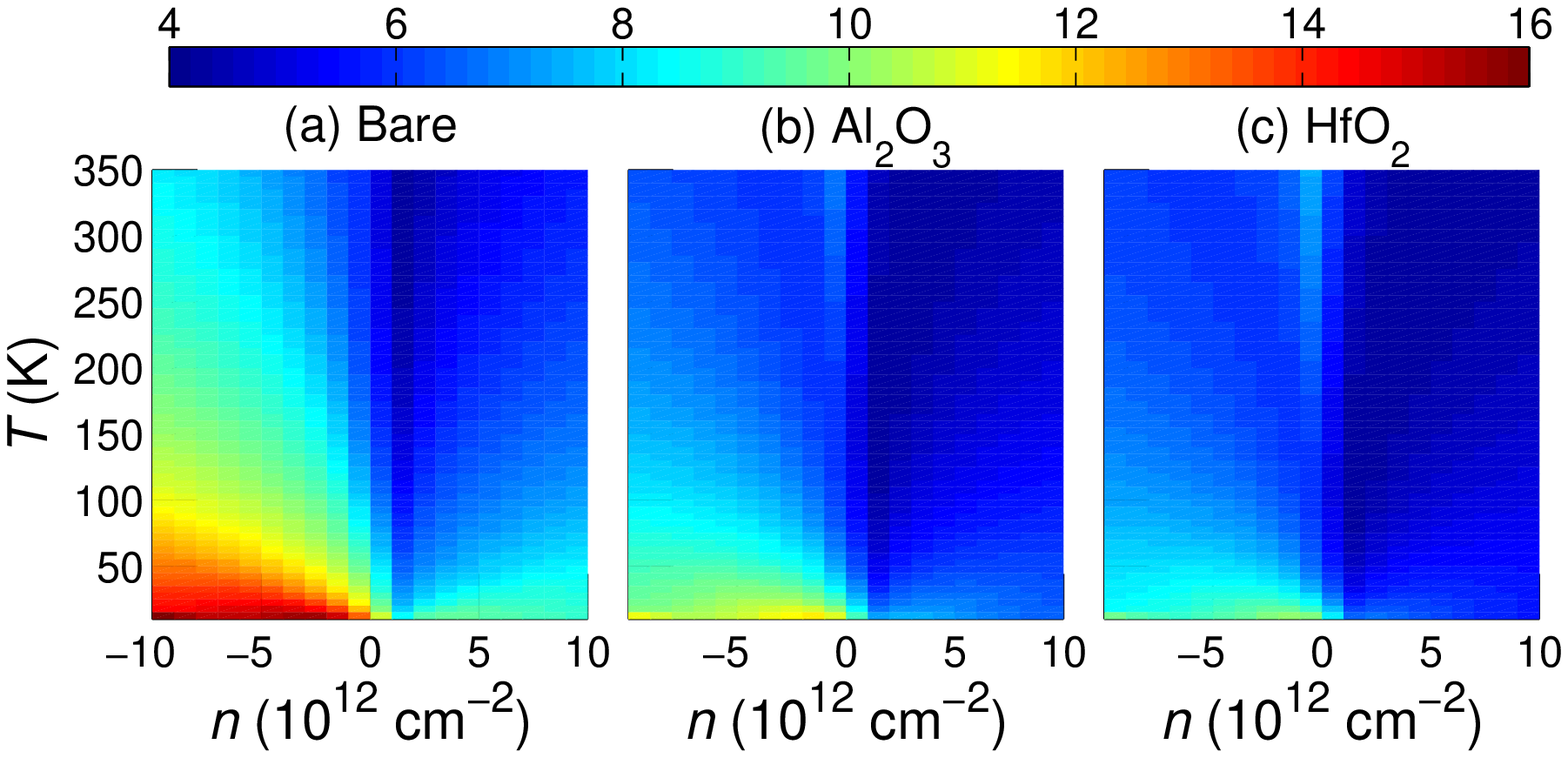}

\caption{We plot the mobility anisotropy factor (defined as $\mu_{\text{AC}}/\mu_{\text{ZZ}}$)
for electrons/holes in phosphorene \textbf{(a)} without, \textbf{(b)}
with Al$_{2}$O$_{3}$ and \textbf{(c)} with HfO$_{2}$ encapsulation,
for $n=-10^{13}$ to $10^{13}$ cm$^{-2}$ from $T=10$ to 350 K.
The electron-doped (hole-doped) region corresponds to positive (negative)
$n$. The mobility anisotropy is smaller for electrons than for holes
because of the larger electron $\mu_{\text{ZZ}}$. The anisotropy
also decreases when we use a high-$\kappa$ top gate oxide or at low
carrier densities or high temperatures. This is because polarization
charge screening of the impurities increases the mobility anisotropy.}

\label{Fig:MobilityAnisotropy}
\end{figure}

Another consequence of the weaker polarization charge screening is
smaller mobility anisotropy. We plot the anisotropy factor $\mu_{\text{AC}}/\mu_{\text{ZZ}}$
for electrons and holes in Fig.\ \ref{Fig:MobilityAnisotropy}. In
Fig.\ \ref{Fig:MobilityAnisotropy}(a), we observe that the anisotropy
factor for electrons and holes in unencapsulated phosphorene reaches
a maximum of around 10 and 16, respectively, and that it decreases
when the carrier density is reduced or when the temperature rises,
because a lower carrier density or higher temperature results in weaker
charge screening. The dependence of the anisotropy on charge screening
can also be seen when we use a high-$\kappa$ oxide overlayer. Figure\ \ref{Fig:MobilityAnisotropy}(b)
shows the anisotropy factor for Al$_{2}$O$_{3}$-covered phosphorene.
In contrast to the case for unencapsulated phosphorene, the maximum
anisotropy at low temperatures is respectively around 7 and 11 for
electrons and holes in Al$_{2}$O$_{3}$-covered phosphorene, decreasing
to around 5 and 6 at room temperature. A similar reduction in the
electron and hole anisotropy factor can also be seen in HfO$_{2}$-covered
phosphorene {[}Fig.\ \ref{Fig:MobilityAnisotropy}(c){]}. This shows
that the anisotropy is significantly dependent on charge screening
which varies with temperature, carrier density and dielectric environment.
Polarization charge screening enhances $\mu_{\text{AC}}$ more than
$\mu_{\text{ZZ}}$ because the momentum change from CI scattering
for carriers moving in the armchair direction is smaller and more
heavily screened than in the zigzag direction. Thus, when phosphorene
is encapsulated with a high-$\kappa$ overlayer, the effect of polarization
charge screening is reduced, and the increase in $\mu_{\text{ZZ}}$
is proportionately larger than the increase in $\mu_{\text{AC}}$.

\section*{Summary}

In summary, we have studied the CI-limited mobility ($\mu_{\text{AC}}$
and $\mu_{\text{ZZ}}$) for electrons and holes in bare and encapsulated
monolayer phosphorene. We find that the mobility is highly dependent
on the crystal orientation with $\mu_{\text{AC}}$ around an order
of magnitude larger than $\mu_{\text{ZZ}}$ because of the smaller
effective electron and hole masses in the armchair direction. The
mobility in bare monolayer phosphorene decreases with increasing temperature
and decreasing carrier density because of the weaker polarization
charge screening of impurities. The same trend can be seen for the
mobility anisotropy factor ($\mu_{\text{AC}}/\mu_{\text{ZZ}}$) since
the reduced polarization charge screening has a greater negative effect
on charge transport in the armchair direction. 

We also find that the mobility is more temperature-dependent in the
armchair direction than the zigzag direction. When a high-$\kappa$
overlayer is used to encapsulate the phosphorene, the carrier mobility
is enhanced by up to an order of magnitude because of dielectric screening
of the charged impurities. The mobility enhancement from encapsulation
is also relatively larger for $\mu_{\text{ZZ}}$ than for $\mu_{\text{AC}}$.
The dependence of the mobility on carrier density and temperature
is reduced by encapsulation because of the weaker polarization charge
screening. This also results in smaller mobility anisotropy, which
should be detectable in experiments and may be helpful for reducing
the orientation dependence in charge transport. Our results suggest
that encapsulation is an effective strategy for improving charge transport
in monolayer phosphorene. 

The authors gratefully acknowledge the financial support from the
Agency for Science, Technology and Research (A{*}STAR), Singapore.

\bibliographystyle{apsrev4-1}
\bibliography{references}

\end{document}